\documentclass[12pt,a4paper]{article}
\usepackage[a4paper, total={6in,9in}]{geometry}

\usepackage{amsmath}
\usepackage{amsfonts}
\usepackage{amssymb}
\usepackage{graphicx}
\usepackage{color}

\usepackage{xcolor,colortbl}
\usepackage[thinlines]{easytable}
\usepackage{caption}
\usepackage{placeins}
\usepackage{color,soul}

\usepackage{lineno}

 \usepackage{ulem}

\title{Atomic Force Nanoscope}

\author
{\normalsize{Alessandro Siria,$^{1}$ Antoine Nigu\`{e}s$^{1}$} \\
\small{$^{1}$Laboratoire de Physique Statistique, Ecole Normale Sup\'{e}rieure, }\\
\small{UMR CNRS 8550, PSL Research University,}\\
\small{75005 Paris Cedex 05, France}\\
\small{$^\star$ alessandro.siria@ens.fr, antoine.nigues@ens.fr, }
}

\begin{document}

\baselineskip24pt

\maketitle
\begin{abstract}
\linespread{1.6}

Atomic Force Microscopy (AFM) allows to probe matter at atomic scale by measuring the perturbation of a nanomechanical oscillator induced by near-field interaction forces. 
The quest to improve sensitivity and resolution of AFM has forced the introduction of a new class of  resonators with dimensions well below the micrometer scale.

In this context, nanotube resonators are the ultimate mechanical oscillators because of their one dimensional nature, small mass and almost perfect crystallinity,
coupled to the possibility of functionalisation 
, these properties make them the perfect candidates as ultra sensitive, on-demand force sensors. However their tiny dimensions make the measurement of the mechanical properties a very challenging task in particular when working in cavity free geometry at ambient temperature. By using a highly focused electron beam, we show that the mechanical response of nanotubes can be quantitatively measured while approaching to a surface sample. By coupling electron beam detection of individual nanotubes with a custom AFM  we can image the surface topography of a sample by measuring in real time the mechanical properties of the nanoresonators. The combination of very small size and mass together with the high resolution of the electron beam detection method offers unprecedented opportunities for the development of a new class of Atomic Force Nanoscope (AFN).

\end{abstract}

\maketitle

Nanoscience and nanotechnology rely on the ability to manipulate and probe objects with a resolution in the deep nanometer range [1,2,3,4]. In the last three decades, advances in the field have been possible mainly thanks to the development of Atomic Force Microscopy (AFM) [5,6,7]. The idea behind the technique is at the same time very simple to understand and impressive in the attainable results: a tiny mechanical oscillator with a sharp tip at the extremity is approached to a surface and the evolution of the mechanical properties of the probe is monitored in real time to get information about the sample topography and properties. Generally the mechanical oscillator used in AFM is a micrometer sized silicon cantilever presenting a very sharp tip at the extremity [8]. The need to increase the force sensitivity and spatial resolution, pushed the researchers to develop alternative mechanical oscillators based on self-assembled structures with submicrometer dimensions. Prominent examples are silicon based nanowires [9,10] and carbon or boron nitride nanotubes [11,12,13,14] and even bidimensional suspended membranes [15,16]. These new probes, because of their minute size and mass and almost perfect cristallinity, couple high spatial resolution and force sensitivity reaching the zepto-newton ($10^{-24}$ N) range [17] for carbon nanotubes working in cryogenic environment. Despite their potentiality for Scanning Probe Micriscopy, so far nanotubes have never been used in microscopy paractical applications. 
The major drawback for the further development of this new kind of probes is the extreme difficulty to detect their position and motion. It is worth mentioning that the mechanical properties of suspended nanowires have been elegantly measured via optical interferometry [18] leading to a first proof of concept of nanowire based scanning force microscope [19]. Despite the effort and recent progress, optical detection schemes can be hardly applied to nanotubes which are the ultimate mechanical oscillators and force sensors: their diameter, in fact, ranging from 1 nm to few tens of nanometers, is too small to be possibly detected with optical techniques. Very recently, electron microscopy has been shown to detect the thermally induced resonant properties of nanomechanical resonators with picogram effective masses [9] and even attogram scale carbon nanotubes [20]. However, to the best of our knowledge, this technique has not been applied for nanotube based force microscopy.\\
In this study we demonstrate that electron beam detection of mechanical oscillators can be harnessed to develop the first example of scanning force microscope based on individual suspended nanotubes. The individual nanotube is mounted on top of a custom AFM working {\it in situ} in a Scanning Electron Microscope (SEM) at a pressure of $\approx 10^{-5}$ mbar. The highly focused electron beam of the SEM is positioned, in the so-called "spot mode", on the externally driven nano-oscillator and the inelastically scattered electrons (SE) are monitored to measure the resonator motion with a spatial resolution of 0.4 nm. As in standard non-contact AFM, the mechanical response of the nanotube is monitored in real time via amplitude and phase locking, while approaching a sample, allowing to reconstruct the surface topography. This new technique allows to take advantage of the exceptional properties of nanotubes as force sensors for scanning microscopy. For standard electron microscopy, the spatial resolution, given by the electron beam size, is below 1 nm and it represents a three orders of magnitude improvement with respect to optical techniques. The high focusing of the electron beam further increases the motion detection resolution for unidimensional systems[9,20]. We therefore demonstrate for the first time, that one dimensional oscillators can be really harnessed and used for scanning force microscopy.\\
In this work we used boron nitride multiwalled nanotubes (BNNTs) realized via Chemical Vapour Deposition (CVD) [21]. 
CVD BNNTs present a very high structural purity with no evident defects in a spatial region reaching the micrometric range. The nanotubes are glued at the extremity of an electrochemically etched tungsten tip, fixed on a three axis piezo inertial motor mounted inside a SEM (FEI-Nova nanoSEM). In figure 1 we present a scheme of the experimental set-up and a SEM picture of one nanotube used in this study in front of an electrochemically etched tungsten tip. The electron beam is set fixed at a chosen working point along the section of the NT. This working point is chosen so that the variation of the SE intensity around this working point stays linear, see figure 1e. To counteract the drift of the nanotube with respect to the electron beam, a feedback loop acts directly on the SEM deflection coils to control the position of the electron beam and keep constant the low frequency component of emitted secondary electron (SE) intensity detected via a Everhart-Thornley Detector (ETD).This method is a significant breaktrough in the field of position control and act like a 1D nano-GPS. This allows to maintain, for the whole duration of the experiment, the electron beam at a constant working position. The intensity of the emitted SE are recorded in real time via a Specs Nanonis SPM electronics and analyzed to extract the DC and AC fluctuating components.

\begin{figure}[!htb]
\centering
   \includegraphics[width=\columnwidth]{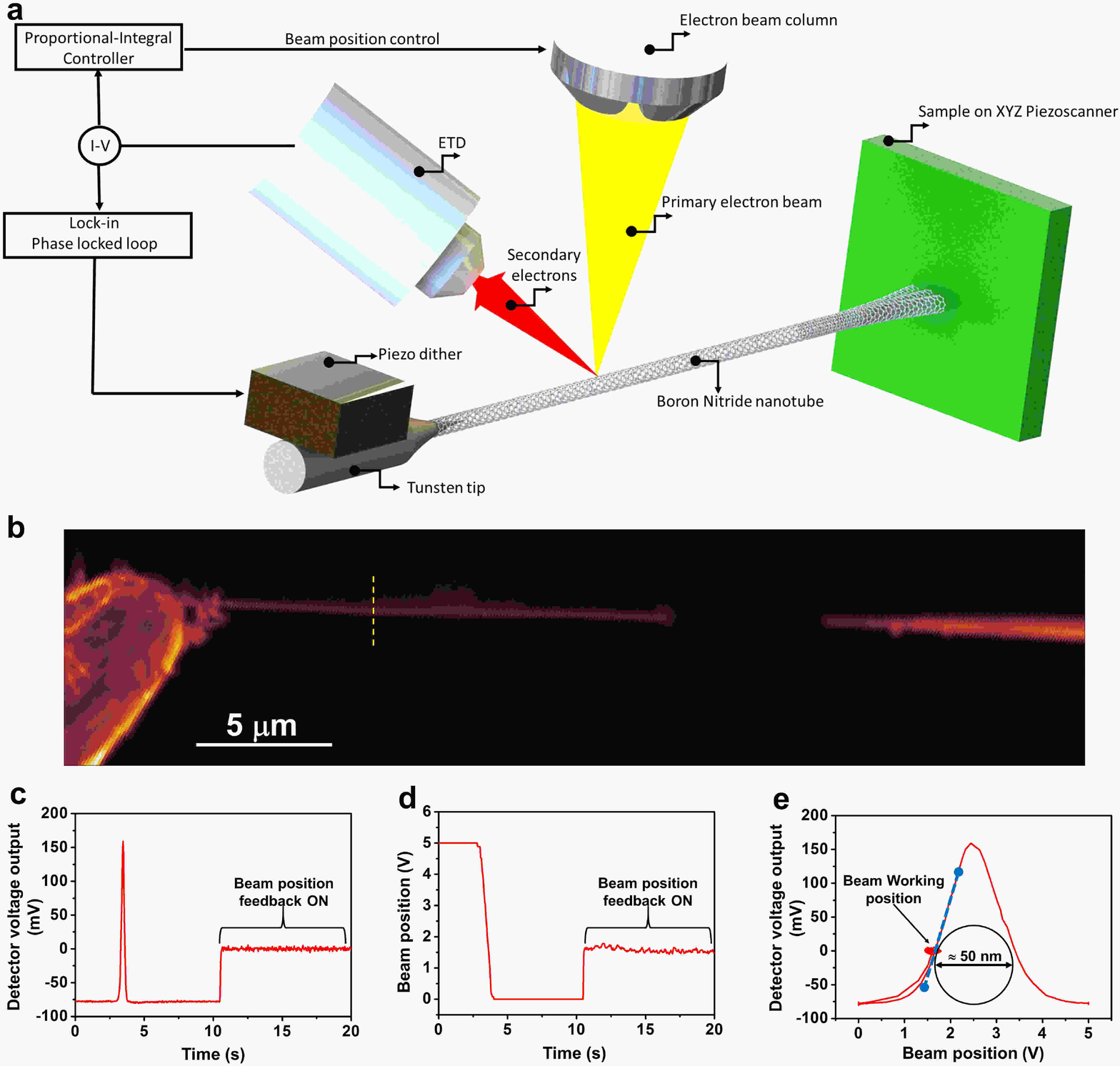}
      \caption{\label{figure1}\textbf{Experimental set-up. a)} Schematic drawing of the experimental set-up; a boron nitride nanotube on top of a custom AFM is excited at the resonant frequency by an external piezo dither. The secondary electrons emitted by the nanotube irradiated by an electron beam are recorded via an ETD detector. The DC component of the signal is used to fix the position of the electron beam with respect to the nanotube by acting on the electron beam deflection coils through a feedback loop; the AC intensity fluctuations record the mechanical motion of the oscillator. {\bf b)} Scanning Electron Microscope picture of one boron nitride nanotube used during the study; measured nanotube diameter $D=50$ nm and length $L=18\ \mu$m in front of a tungsten tip. Yellow dashed line corresponds to the cross section presented in c) and e). {\bf c)} Secondary electron intensity converted in voltage detected during a line scan over the nanotube; at $t=10$ s the feedback loop acting on the deflection coils is activated to keep the SE intensity at a defined set-point, 0 V here. {\bf c)} Voltage applied to the deflection coils to control the position of the electron beam;  at $t=10$ s the feedback loop is activated to keep constant the low frequency component of SE intensity. {\bf e)} SE intensity (b) as a function of the voltage applied (c) to control the position of the electron beam; once the feedback loop is activated the working position (red sputtered) is attained and the intensity of the SE isfixed by changing the applied voltage to the deflection coils; the slope of the blue dashed line determines the amplitude calibration to be 125 nm/V for this position along the tube}
\end{figure}

BN nanotubes are mechanically excited with an external piezo dither and the variations of SE, related to the relative displacement of the NT around the working point, are recorded as a function of the excitation frequency. When excited by an external sinusoidal force $F_{ext}(\omega)=F_{ext}e^{i\omega t}$, the nanotube behaves in first approximation as a spring-mass system with oscillation amplitude and phase with respect to the excitation given by:
\begin{eqnarray}
A(\omega)=\frac{F_{ext}}{\sqrt{m^2_{eff}(\omega^{2}_{0}-\omega^{2})^2+\gamma^2\omega^{2}}} \\
\phi(\omega)=\arctan\left(\frac{\gamma\omega}{m_{eff}(\omega^{2}_{0}-\omega^{2})}\right)
\end{eqnarray}
with $m_{eff}$ the oscillator effective mass, $\omega_{0}=\sqrt{\frac{k}{m_{eff}}}$ the resonant frequency and $\gamma$ the damping factor. In figure 2a we present the response of a nanotube with diameter $D\approx$50 nm and length $L=18\ \mu$m, when varying the frequency of excitation. The mechanical response is characterised by a resonance at $f_{0}=\omega_{0}/2\pi\approx 270$  kHz and a quality factor, defining the inverse of the damping rate, $Q\approx 250$. 
In contrast to previous studies on nanowires and single walled carbon nanotubes [19,9,20], we do not observe a double resonant peak due to non degeneracy of polarisations. This further confirms the high structural quality and geometrical symmetry of BN nanotubes used during this work.\\
\begin{figure}[!htb]
\centering
   \includegraphics[width=\columnwidth]{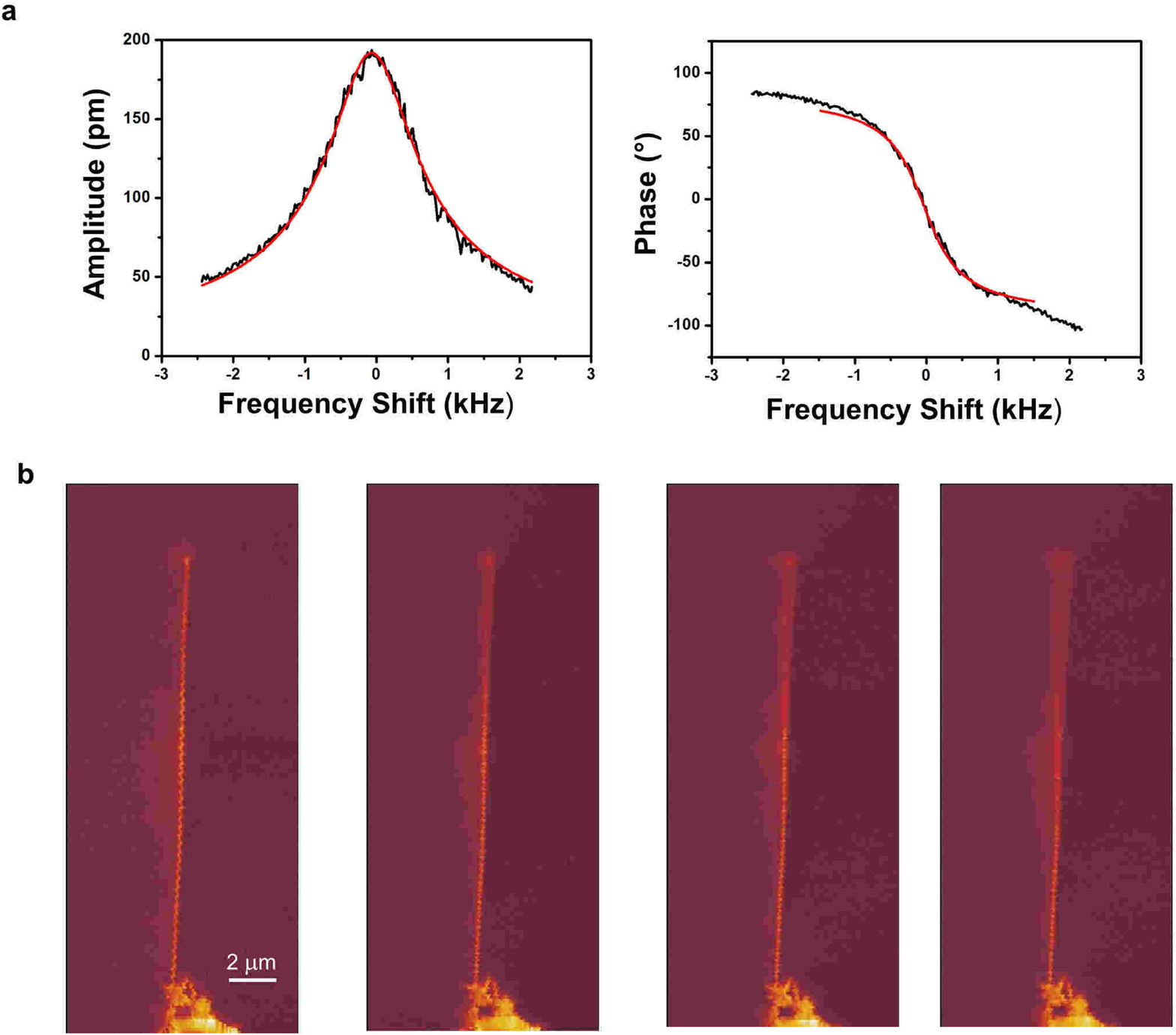}
       \caption{\textbf{Electron beam detection of the mechanical response of a BN nanotube. a)} Amplitude and phase of the nanotube mechanical response as a function of the exciting frequency. The frequency is scanned around the resonant frequency; for a BN nanotube with $D=50$ nm and $L=18\ \mu$m the resonance is measured at $f_{0}=\omega_{0}/2\pi\approx 270$ and the quality factor $Q\approx 250$. Red curves are best fit based on equations (1) and (2). {\bf b)} Scanning Electron Microscope pictures of the nanotube for different excitation amplitudes: from left to right 0 mV, 500 mV, 1 V, 2 V; the mode shape reproduces well the standard Euler-Bernoulli profile for the fundamental mode of a single clamped tube.}
        \label{figure2}
\end{figure}
To confirm that the detected resonance is due to the mechanical motion of the nanotube, we switch the SEM operation mode from "spot mode"  to the conventional full frame imaging: when excited at the resonance the nanotube position exhibits the standard resonant profile for the fundamental mode of a single clamped oscillator, as described by the Euler-Bernoulli equation. In figure 2c  we plot the SEM images for different excitation amplitudes confirming that for the range of excitations used during the experiment, the oscillator stays in its linear regime.\\
As the interaction of the oscillator with its environment is modified, one observes a change in both the frequency and the amplitude at resonance. The shift in resonance frequency $\delta f$ is related to the conservative force response, whereas the broadening of the resonance (change of quality factor $Q_0\rightarrow Q_1$) is related to dissipation [22]:
\begin{eqnarray}
\frac{\partial F_i}{\partial r_i}= 2 k_i \frac{\delta f_i}{f_{0,i}} \text{ and } F_{D,i} = \frac{k_i}{\sqrt{3}} \left( \frac{1}{Q_0} -\frac{1}{Q_1} \right)
\end{eqnarray}
where the index $i$, indicates the component of the position vector. In practice, during a typical experiment, two feedback loops allow us to work at the resonance and maintain constant the oscillation amplitude $a_0$. Monitoring the frequency shift $\delta f$ and the resonance quality factor  $Q_i$, obtained from the excitation voltage or the oscillation amplitude, thus provides a direct measurement of real and imaginary parts of the mechanical impedance.\\
To demonstrate the possibility to use a nanotube as the force sensor of a scanning probe microscope, we firstly approached a very sharp tungsten tip to the oscillating nanotube. While approaching the tungsten tip down to the contact we record the evolution of the phase between the nanotube oscillation and the exciting dither and of the oscillation amplitude. This first experiment is performed in an open-loop configuration: any variation of the interacting environment is directly measured by a change of both the oscillation amplitude and the phase with respect to the excitation. Once the mechanical response of the NT is modified by the tip-NT interaction, we kept the distance constant and we performed a scanning picture of the tungsten tip, as in standard AFM. As shown in figure 3a, the amplitude and phase shift as a function of the position of the tip correctly detect the spatial force field induced by the tungsten, and consequently reconstruct the tip shape. \\
\begin{figure}[!htb]
\centering
   \includegraphics[width=1\columnwidth]{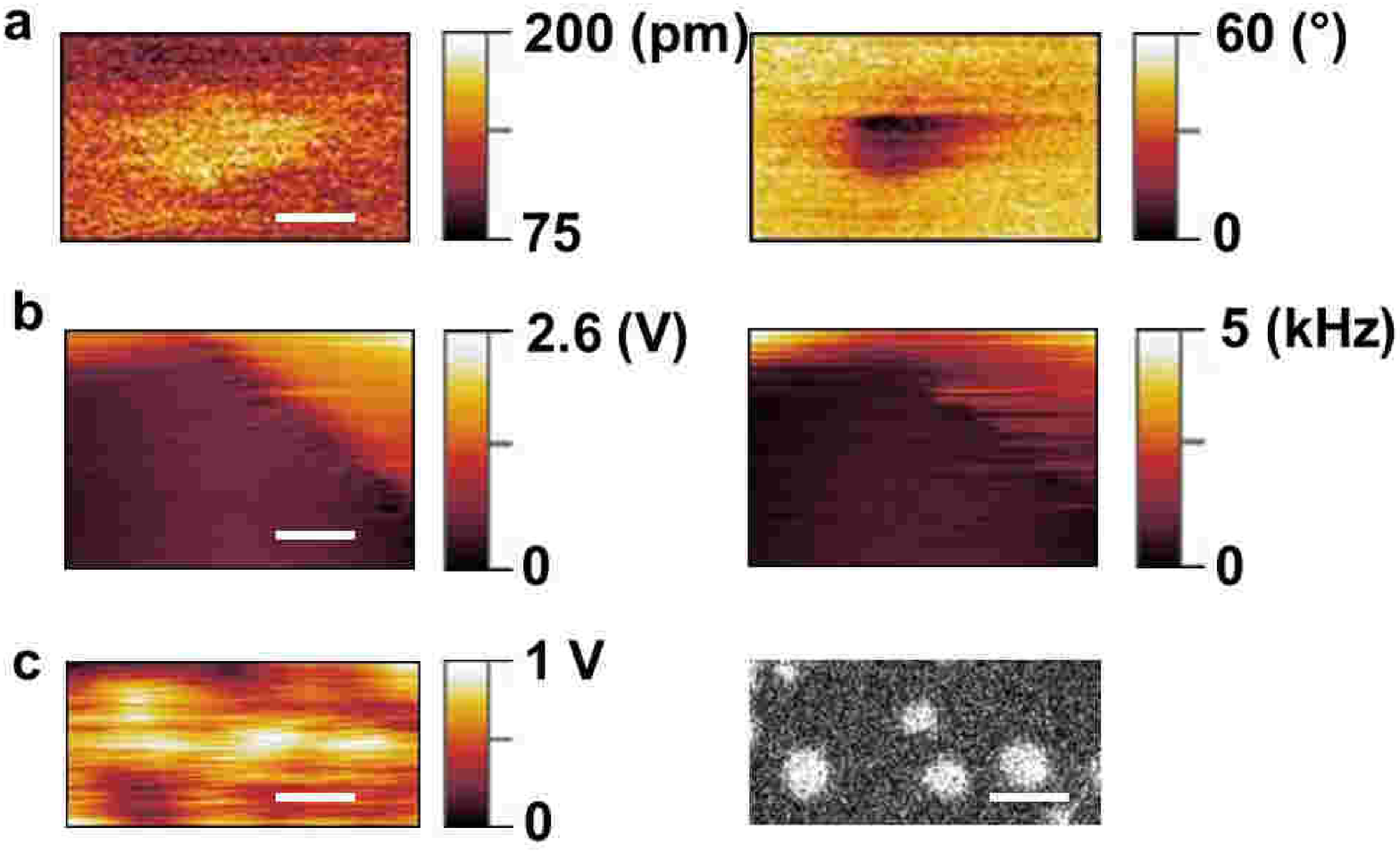}
       \caption{\textbf{Nanotube-based Scanning Force Microscopy. a)} Amplitude (left) and phase (right) of the nanotube mechanical motion while scanning a sharp tungsten tip; the measurement reproduces correctly the force field induced by the tip and reconstructs the tip shape. Scale bar is 200 nm. {\bf b)} Scanning force imaging of a (100) silicon surface; the excitation amplitude keeping the nanotube oscillation amplitude constant (left) and the frequency shift (right) reproduce the surface topography of a 1 $\mu$m terrace. Scale bar is 500 nm. {\bf c)} Static AFM image (left) and SEM picture(right) of gold nanoparticles deposited on a HOPG sample. The static AFM image is obtained by recording the deflection of the nanotube due to its interaction with the substrate; this quantity is given by the voltage command to the deflection coils which keeps constant the SE emission intensity. Scale bar is 500 nm}
        \label{figure3}
\end{figure}
As a further demonstration of the potentialities of the NT-SFM, a BNNT (diameter $D=80$ nm and length $L=20\ \mu$m) with resonant frequency $f_0\approx350$ kHz and oscillation amplitude of $\approx$ 10 nm scans at a distance of $\approx$ 10 nm  a (100) silicon surface while a double feedback loop acts on both the phase and the amplitude. In figure 3b we plot a cartography of a 1 micron silicon terrace, recording at the same time the excitation voltage applied to keep constant the amplitude and the frequency shift. Variations of both quantities allow to reconstruct the surface topography with a resolution of $\approx$ 100 nm, given by the size of the nanotube apex.\\
Eventually we take benefit from the versatility of the NT-SFM and in addition to what is presented above, we show in figure 3c, a static AFM and a SEM picture of gold nanoparticles deposited on a HOPG sample. In this case, by static we mean that the nanotube is neither excited nor oscillating: this operation mode is the analog of the static mode employed in standard Atomic Force Microscopy. The static AFM picture is obtained by recording the deflection of the nanotube during the scan; experimentally this quantity is given by the voltage command to the deflection coils to keep constant the SE emission intensity. When the nanotube deflects because of the interaction with the surface, its relative position with respect to the electron beam changes, therefore modifying the SE emission intensity. The feedback loop acting on the beam position allows to reconstruct precisely the displacement of the nanotube and consequently the surface topography.\\
In addition, it is worth mentioning the complementarity of this operation mode with the AC mode presented before: by recording the deflection of the nanotube it is in fact possible to quantitatively measure the total force which the nanotube is submitted to. Together with the measurement of the conservative force gradient via the frequency shift and the dissipative forces via the change in the quality factor, this method allows to completely characterise the force environment with an unprecedented resolution.\\
Altogether our pioneering work demonstrates that one dimensional oscillators such as boron nitride nanotubes can be harnessed as probe sensors for Scanning Force Microscopy. By using a highly focused electron beam we can detect and record the mechanical response of an individual nanotube. Coupling atomic force microscopy technology together with high resolution electron beam detection, allows to measure, in real-time, conservative and dissipative interactions as well as the complete static interactions of a nanotube with its environment. We elegantly exploited this new detection method to measure the surface topography of samples approached by the extremity of an individual nanotube.
More in general, the highly focused electron beam allows to detect oscillators with diameters in the nanometer range, drastically increasing the force sensitivity and spatial resolution compared to standard atomic force microscope cantilevers and even more recent nanowires. Despite the current low quality of the images obtained in this first work, the technique here presented represents a clear breakthorugh for the field and pave the way for the development of a new class of ultrasensitive nanotube-based scanning force microscopes.


\noindent{\bf Acknowledgements} \\
Authors acknowledge funding from the European Union's H2020 Framework Programme / ERC Starting Grant agreement number 637748 - NanoSOFT.

\small

\section*{References}

\textcolor[rgb]{1,1,0.88}{hviyviyviyv}

[1] A. Nigu\`{e}s, A. Siria, P. Vincent, P. Poncharal and L. Bocquet, Ultra-high interlayer friction inside Boron-Nitride nanotubes, Nature Materials 13 688-693 (2014).

[2] Ternes, M., Lutz, C. P., Hirjibehedin, C. F., Giessibl, F. J., Heinrich, A. J. (2008). The Force Needed to Move an Atom on a Surface Measuring the Charge State of an Adatom with Noncontact Atomic Force Microscopy, 1066(111).

[3] Socoliuc, A. (2006). Atomic Scale control of Friction by actuation of nanometer-sized contacts. Science, 313(207), 20710. https://doi.org/10.1126/science.1125874

[4] Kisiel, M., Gnecco, E., Gysin, U., Marot, L., Rast, S., Meyer, E. (2011). Suppression of electronic friction on Nb ﬁlms in the superconducting state. Nature Materials, 10(2), 119122. https://doi.org/10.1038/nmat2936

[5] Binnig, G., Quate, C. F., Gerber, C. (1986). Atomic force microscope. Phys. Rev. Lett, 56, 930934. https://doi.org/10.1103/PhysRevLett.67.1582

[6] Giessibl, F. J. (1995). Surface by Atomic Force Microscopy. Science, 267(13), 68. https://doi.org/10.1126/science.267.5194.68

[7] Giessibl, F. J. (2003). Advances in atomic force microscopy. Reviews of Modern Physics, 75(3), 949983. https://doi.org/10.1103/RevModPhys.75.949

[8] Barwich, V., Bammerlin, M., Baratoﬀ, A., Bennewitz, R., Guggisberg, M., Loppacher, C., Forr, L. (2000). Carbon nanotubes as tips in non-contact SFM. Applied Surface Science, 157(4), 269273. https://doi.org/10.1016/S0169-4332(99)00538-3

[9] A. Nigu\`{e}s, A., Siria, A., Verlot, P. (2015). Dynamical Backaction Cooling with Free Electrons. Nature Communications, 8. https://doi.org/10.1038/ncomms9104

[10] Gloppe, A., Verlot, P., Dupont-Ferrier, E., Kuhn, A. G., Siria, A., Poncharal, P., Arcizet, O. (2014). Bidimensional nano-optomechanics and topological backaction in a nonconservative radiation force ﬁeld. International Conference on Optical MEMS and Nanophotonics, (September), 7374. https://doi.org/10.1109/OMN.2014.6924599

[11] Poncharal, P., Wang, Z., Ugarte, D., de Heer WA. (1999). Electrostatic deﬂections and electromechanical resonances of carbon nanotubes. Science (New York, N.Y.), 283(5407), 15136. Retrieved from http://www.ncbi.nlm.nih.gov/pubmed/10066169

[12] Moser, J., Eichler, A., Gttinger, J., Dykman, M. I., Bachtold, A. (2014). Nanotube mechanical resonators with quality factors of up to 5 million. Nature Nanotechnology, 9(12), 10071011. https://doi.org/10.1038/nnano.2014.234

[13] Steele, G. A. (2011). Strong Coupling Between Single-Electron Tunneling and Nanomechanical Motion. Science, 1103(2009). https://doi.org/10.1126/science.1176076

[14] Lassagne, B., Tarakanov, Y., Kinaret, J., Garcia-Sanchez, D., Bachtold, A. (2009). Coupling mechanics to charge transport in carbon nanotube mechanical resonators. Science (New York, N.Y.), 325(5944), 11071110. https://doi.org/10.1126/science.1174290

[15] Miao, T., Yeom, S., Wang, P., Standley, B., Bockrath, M. (2014). Graphene nanoelectromechanical systems as stochastic-frequency oscillators. Nano Letters, 14(6), 29822987. https://doi.org/10.1021/nl403936a

[16] De Alba, R., Massel, F., Storch, I. R., Abhilash, T. S., Hui, A., McEuen, P. L., Parpia, J. M. (2016). Tunable phonon cavity coupling in graphene membranes. Nature Nanotechnology, (June). https://doi.org/10.1038/nnano.2016.86

[17] Chaste, J., Eichler, A., Moser, J., Ceballos, G., Rurali, R., Bachtold, A. (2012). A nanomechanical mass sensor with yoctogram resolution. Nature Nanotechnology, 7(5), 301304. https://doi.org/10.1038/nnano.2012.42

[18] de Lpinay, L. M., Pigeau, B., Besga, B., Vincent, P., Poncharal, P., Arcizet, O. (2016). A universal and ultrasensitive vectorial nanomechanical sensor for imaging 2D force ﬁelds. Nature Nanotechnology, (October), 17. https://doi.org/10.1038/nnano.2016.193

[19] Rossi, N., Braakman, F. R., Cadeddu, D., Vasyukov, D., Ttncoglu, G., Morral, A. F. i, Poggio, M. (2016). Vectorial scanning force microscopy using a nanowire sensor. Nature Nanotechnology, (October), 110. https://doi.org/10.1038/nnano.2016.189

[20] Tsioutsios, I., Tavernarakis, A., Osmond, J., Verlot, P., Bachtold, A. (2017). Real-Time Measurement of Nanotube Resonator Fluctuations in an Electron Microscope. Nano Letters, 17(3), 17481755. https://doi.org/10.1021/acs.nanolett.6b05065

[21] Bechelany, M., Brioude, A., Stadelmann, P., Bernard, S., Cornu, D., Miele, P. (2008). Preparation of BN Microtubes / Nanotubes with a Unique Chemical Process. The Journal of Physical Chemistry C, 112(47), 1832518330.

[22] Karrai, K., Grober, R. D. (1995). Piezoelectric tip-sample distance control for near ﬁeld optical microscopes. Applied Physics Letters, 66(14), 1842. https://doi.org/10.1063/1.113340


\end{document}